\documentclass[prl,twocolumn,letterpaper,superscriptaddress,reprint]{revtex4-1}

\usepackage{graphicx}
\usepackage{pslatex}
\usepackage{amsmath}
\usepackage{booktabs, caption, makecell}
\usepackage{threeparttable}

\newcommand{\atUCLA}{Dept. of Physics and Astronomy, Univ. of California, Los Angeles, Los Angeles, CA 90095.}
\newcommand{\atOSU}{Dept. of Physics, Ohio State Univ., Columbus, OH 43210.}
\newcommand{\atUH}{Dept. of Physics and Astronomy, Univ. of Hawaii, Manoa, HI 96822.}
\newcommand{\atNTU}{Dept. of Physics, Grad. Inst. of Astrophys.,\& Leung Center for Cosmology and Particle Astrophysics, National Taiwan University, Taipei, Taiwan.}
\newcommand{\atUCI}{Dept. of Physics, Univ. of California, Irvine, CA 92697.}
\newcommand{\atKU}{Dept. of Physics and Astronomy, Univ. of Kansas, Lawrence, KS 66045.}
\newcommand{\atWashU}{Dept. of Physics, Washington Univ. in St. Louis, MO 63130.}
\newcommand{\atSLAC}{SLAC National Accelerator Laboratory, Menlo Park, CA, 94025.}
\newcommand{\atUD}{Dept. of Physics, Univ. of Delaware, Newark, DE 19716.}
\newcommand{\atUCL}{Dept. of Physics and Astronomy, University College London, London, United Kingdom.}
\newcommand{\atUMinn}{School of Physics and Astronomy, Univ. of Minnesota, Minneapolis, MN 55455.}
\newcommand{\atJPL}{Jet Propulsion Laboratory, Pasadena, CA 91109.}
\newcommand{\atCCAP}{Center for Cosmology and Particle Astrophysics, Ohio State Univ., Columbus, OH 43210.}
\newcommand{\atChicago}{Dept. of Physics, Enrico Fermi Institute, Kavli Institute for Cosmological Physics, Univ. of Chicago , Chicago IL 60637.}

\newcommand{\atCalPoly}{Physics Dept., California Polytechnic State Univ., San Luis Obispo, CA 93407.}
\begin{document}

\title{Characteristics of Four Upward-pointing Cosmic-ray-like Events Observed with ANITA}

\author{P.~W.~Gorham}
\affiliation{\atUH}

\author{J.~Nam}
\affiliation{\atNTU}

\author{A.~Romero-Wolf}
\affiliation{\atJPL}

\author{S.~Hoover}
\affiliation{\atUCLA}

\author{P.~Allison}
\affiliation{\atOSU}
\affiliation{\atCCAP}

\author{O.~Banerjee}
\affiliation{\atOSU}

\author{J.~J.~Beatty}
\affiliation{\atOSU}
\affiliation{\atCCAP}

\author{K.~Belov}
\affiliation{\atJPL}

\author{D.~Z.~Besson}
\affiliation{\atKU}

\author{W.~R.~Binns}
\affiliation{\atWashU}

\author{V.~Bugaev}
\affiliation{\atWashU}

\author{P.~Cao}
\affiliation{\atUD}

\author{C.~Chen}
\affiliation{\atNTU}

\author{P.~Chen}
\affiliation{\atNTU}

\author{J.~M.~Clem}
\affiliation{\atUD}

\author{A.~Connolly}
\affiliation{\atOSU}
\affiliation{\atCCAP}

\author{B.~Dailey}
\affiliation{\atOSU}

\author{C.~Deaconu}
\affiliation{\atChicago}

\author{L.~Cremonesi}
\affiliation{\atUCL}

\author{P.~F.~Dowkontt}
\affiliation{\atUCLA}

\author{M.~A.~DuVernois}
\affiliation{\atUH}

\author{R.~C.~Field}
\affiliation{\atSLAC}

\author{B.~D.~Fox}
\affiliation{\atUH}

\author{D.~Goldstein}
\affiliation{\atUCI}

\author{J.~Gordon}
\affiliation{\atOSU}

\author{C.~Hast}
\affiliation{\atSLAC}

\author{C.~L.~Hebert}
\affiliation{\atUH}

\author{B.~Hill}
\affiliation{\atUH}

\author{K.~Hughes}
\affiliation{\atOSU}

\author{R.~Hupe}
\affiliation{\atOSU}

\author{M.~H.~Israel}
\affiliation{\atWashU}

\author{A.~Javaid}
\affiliation{\atUD}

\author{J.~Kowalski}
\affiliation{\atUH}

\author{J.~Lam}
\affiliation{\atUCLA}

\author{J.~G.~Learned}
\affiliation{\atUH}

\author{K.~M.~Liewer}
\affiliation{\atJPL}

\author{T.C. Liu}
\affiliation{\atNTU}

\author{J.~T.~Link}
\affiliation{\atUH}

\author{E.~Lusczek}
\affiliation{\atUMinn}

\author{S.~Matsuno}
\affiliation{\atUH}

\author{B.~C.~Mercurio}
\affiliation{\atOSU}

\author{C.~Miki}
\affiliation{\atUH}

\author{P.~Mio\v{c}inovi\'c}
\affiliation{\atUH}

\author{M.~Mottram}  
\affiliation{\atUCL}

\author{K.~Mulrey}  
\affiliation{\atUD}

\author{C.~J.~Naudet}
\affiliation{\atJPL}

\author{J.~Ng}
\affiliation{\atSLAC}

\author{R.~J.~Nichol}
\affiliation{\atUCL}

\author{K.~Palladino}
\affiliation{\atOSU}

\author{B.~F.~Rauch}
\affiliation{\atWashU}

\author{K.~Reil}
\affiliation{\atSLAC}

\author{J.~Roberts}
\affiliation{\atUH}

\author{M.~Rosen}
\affiliation{\atUH}

\author{B. Rotter}
\affiliation{\atUH}

\author{J.~Russell}
\affiliation{\atUH}

\author{L.~Ruckman}
\affiliation{\atUH}

\author{D.~Saltzberg}
\affiliation{\atUCLA}

\author{D.~Seckel}
\affiliation{\atUD}

\author{H.~Schoorlemmer}
\affiliation{\atUH}

\author{S.~Stafford}
\affiliation{\atOSU}

\author{J.~Stockham}
\affiliation{\atKU}

\author{M.~Stockham}
\affiliation{\atKU}

\author{B.~Strutt}
\affiliation{\atUCL}

\author{K.~Tatem}
\affiliation{\atUH}

\author{G.~S.~Varner}
\affiliation{\atUH}

\author{A.~G.~Vieregg}
\affiliation{\atChicago}

\author{D.~Walz}
\affiliation{\atSLAC}

\author{S.~A.~Wissel}
\affiliation{\atCalPoly}

\author{F.~Wu}
\affiliation{\atUCLA}
\vspace{2mm}
\noindent

\begin{abstract}
We report on four radio-detected cosmic-ray (CR) or CR-like events observed with the Antarctic Impulsive
Transient Antenna (ANITA), a NASA-sponsored long-duration balloon payload. Two of the four
were previously identified as stratospheric CR air showers during the ANITA-I flight.  A third 
stratospheric CR was detected during
the ANITA-II flight. Here we report on characteristics 
these three unusual CR events, which develop nearly horizontally,  20-30~km above the surface of the Earth.
In addition, we report on a fourth steeply upward-pointing ANITA-I CR-like radio event which has
characteristics consistent with a primary that emerged from the surface of the ice. This suggests a
possible $\tau$-lepton decay as the origin of this event, but such an interpretation would require
significant suppression of the Standard Model ${\tau}$-neutrino cross section.
\end{abstract}
\maketitle

We have previously reported the observation of ultra-high energy (UHE)
cosmic ray (CR) air showers detected from suborbital altitudes with the ANITA balloon payload~\cite{ANITA_CR}
during our first flight in 2007~\cite{HooverThesis}.
The initial blind neutrino-search analysis that led to their identification in the data found 16 events in a signal
box with an expected background of 1.6 events.
Three of these 16 events were deemed background:
two of unknown origin, and one a likely thermal noise fluctuation with no apparent signal content.
The remaining 13 events were consistent with geomagnetically-induced CR radio pulses
seen in reflection off the Antarctic ice surface.
Three additional CRs were also found in cross-correlation analysis after the unblinding,
including two events from directions above the geometric horizon but below the horizontal.
These stratospheric air showers represent a class of CR which has not been previously 
observed.

ANITA~\cite{ANITA-inst} makes precise horizontal (Hpol) and vertical (Vpol) polarization measurements of
each detected impulse, using custom dual-polarized quad-ridged horn antennas.
For the CR events, their nearly horizontal planes of polarization
correlated closely with Lorentz-force components of the predominantly vertical Antarctic 
geomagnetic field, once Fresnel coefficients for reflection
from the ice surface were accounted for. The above-horizon CR events had opposite
polarity compared to the reflected events, consistent with a lack of inversion by reflection,
and also had geomagnetically correlated planes of polarization. In addition to these two above-horizon events observed
in ANITA-I, an additional event of the same type was observed in the 2009 ANITA-II flight, selected according
to its high correlation to CR waveform templates. ANITA-II, which was
optimized for in-ice neutrino detection~\cite{ANITA-II}, did not have
a dedicated CR trigger but still detected a small number of CR impulses that had sufficient signal strength.
Further details of the two flights are given in~\cite{Suppl1}.

\begin{table*}[tb!]
\caption{Expected parameters of the three above-horizon CR events.}
\begin{center}
\begin{threeparttable}
\begin{tabular}{ccccccccccc}
event No. & flight & index & Latitude &Longitude$^{\dag}$ & angle   &   $D^*_{1200}$& $D_{Xmax}$ &   $D_{300}$    & $D_{100}$ &  $H_{X_{max}}$  \\ \hline
5152386 & I & A & 80.2S & 49.0W   &$-4.25\pm0.25^{\circ}$ & $622(+88,-100)$  & $694\pm80$ & $780\pm 77$ & $860\pm 70$ &  $22.0\pm 1.0$  \\
7122397 & I & B & 82.405S &  12.5E   &$-3.4\pm0.32^{\circ}$ & $331(+125,-200)$ & $444(+100,-120)$ & $570\pm 80$ & $667\pm 70$ &  $24.2\pm 2.2$\\
21684774 & II & C & 83.24S &  0.87E   &$-2.3\pm0.3^{\circ}$ & $-83.5(+9,-6)$ & $-17(+189,-75)$ & $285\pm 85$ & $416\pm 70$ & $29.9\pm 1.3$\\
\hline \\
\end{tabular}
\begin{tablenotes}
\item[$^\dag$] Latitude and Longitude of the estimated location of shower maximum $X_{max}$, or for event C, payload location.
\item[*] Distances from payload, in km, to location of indicated shower slant depth in g/cm$^2$.
\end{tablenotes}
\end{threeparttable}
\end{center}
\label{tbl1}
\end{table*}%

Motivated by recent results in which searches for upward-directed or Earth-skimming CR air showers have been  
used to constrain the flux of $\tau$ lepton decays arising from UHE $\nu_{\tau}$~\cite{Auger15, Fargion06, Feng02},
we have performed more detailed evaluation of the properties of these apparently up-coming radio-detected CRs. 
The three stratospheric events appear consistent with our 
expectations for ANITA's acceptance to the known CR flux at energies above
$10^{18}$~eV. In reviewing the other putative background events that passed our blind analysis cuts, we found that
one of these was dominated by Hpol content, consistent with the geomagnetic parameters
of a CR. It arrived at the payload from a direction of  $27.4^{\circ}$ below the horizontal,
which was a fairly typical angle for the reflected CR events.
Yet it did not appear to correlate well with the {\em reflected} CR
signal shape, and was thus rejected as background at the time~\cite{HooverThesis}. 
In re-evaluating this event, we realized that
the polarity and plane of polarization are consistent with an air shower seen directly, without the reflection phase
inversion. However, its steep upward pointing angle poses clear problems for interpretation.
In this report, we analyze characteristics of all four of these unusual upward-directed events seen by ANITA,
with specific focus on what relation, if any,  the previously excluded event may have with $\tau$-lepton-initiated air showers.

 \begin{figure}[htb!]
 \includegraphics[width=3.3in]{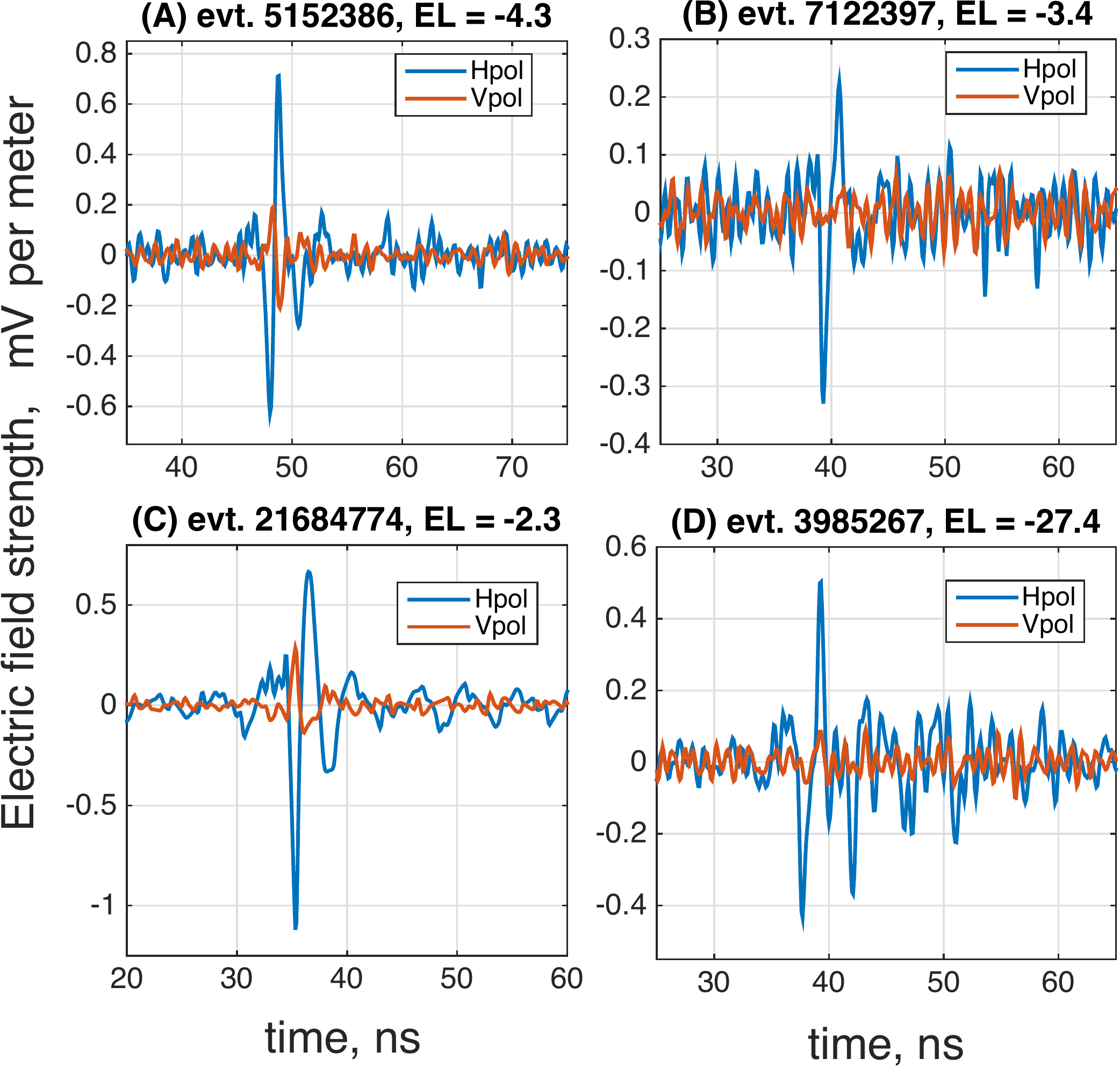}
 \caption{Waveforms for the four events described here. 
 Events are indexed here and in the text by the letters A,B,C,D.
 \label{wfmall}}
 
 \end{figure}

Table~\ref{tbl1} shows characteristics of the three stratospheric events.
Angles of arrival relative to the payload horizontal 
and their standard errors are determined through pulse-phase interferometric mapping~\cite{ARW15}.
Distances to various integrated atmospheric column depths $X$, including the approximate depth of 
shower maximum $X_{max}$, assuming a shower energy of 
$\sim 10^{18}$~eV, are given along the track, based on a standard atmosphere model for
Antarctica, and with uncertainties primarily dominated by the angle-of-arrival uncertainty. 
The geodetic positions in each case are given according to the estimated location of $X_{max}$.

Figure~\ref{wfmall} shows the field-strength waveforms for all of the events, derived
from coherent beam-forming~\cite{ARW15}, with the
instrumental response then deconvolved.  Both Hpol and Vpol are plotted. 
The Hpol polarity of each of these events, checked independently by two quantitative methods,
is phase-reversed
with respect to the other 14 UHECR events which were inverted by reflection from the ice surface~\cite{ANITA_CR}.
For CRs, Vpol polarity and magnitude depends on components of the geomagnetic field in the locale of the event,
as we will quantify later.

The three events at shallow elevation angles, 
which correlate closely in pulse shape to our other sample of radio-detected CRs,
develop and propagate in the stratosphere, under very rarified densities. 
Their overall length is greatly magnified compared to showers observed by ground arrays. The lowest of the three events
has a likely first interaction point well beyond the geometric horizon, and will have largely dissipated in 
the vicinity of the geometric horizon at $\sim 650$~km. The higher two events are at least 200~km, 
and possibly more than 600~km
in length, in both cases passing by the ANITA payload before they have dissipated. In the highest event,
which develops above 30~km, the shower was near its maximum development when it passed by ANITA.
Geometric estimates of ANITA's expected rate of CRs at these angles, using the
acceptance determined by the reflected CRs~\cite{Harm16}, indicates that the number of detected events
is consistent with the known CR spectrum at EeV energies.

 \begin{figure*}[tb!]
 \includegraphics[width=6in]{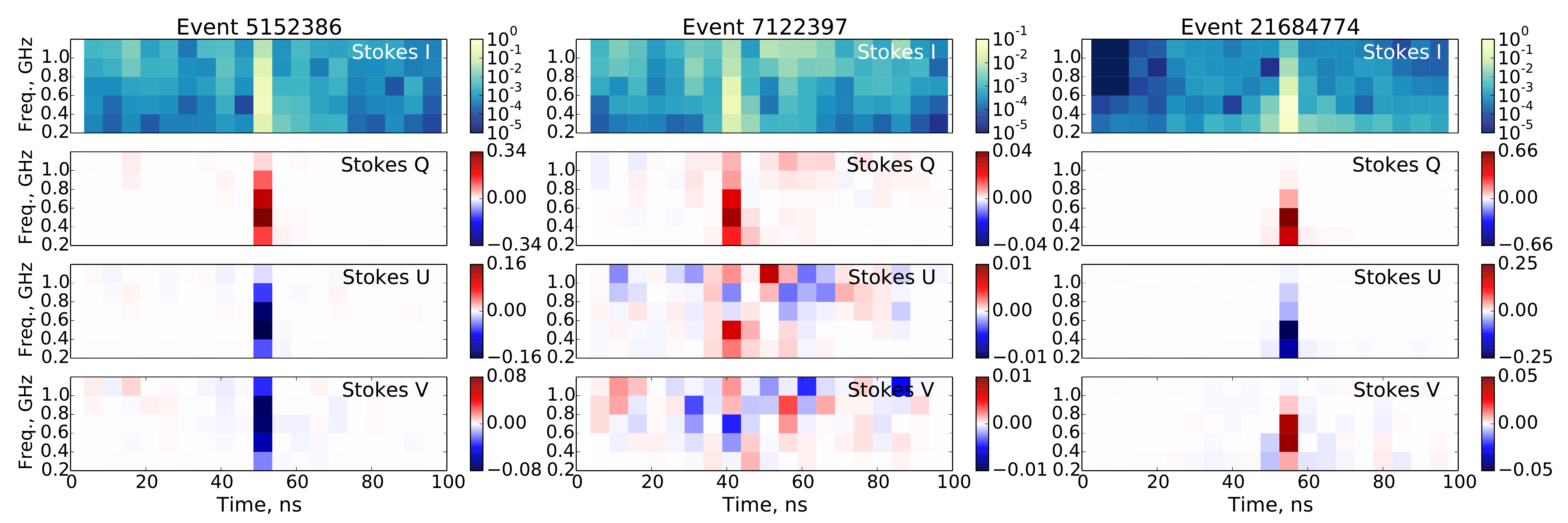}
 \caption{Stokes parameters for the three above-horizon events in the sample considered here.}
 \label{stokesAbove}
 \end{figure*}

To characterize these events more fully, we estimate their Stokes parameters. Fig.~\ref{stokesAbove} show
I,Q,U,V in a spectro-temporal decomposition for these three events.
In all cases the linear polarization components associated with Q and U are clearly evident.
In addition, in the two stronger events there is up to 25\% Stokes V content, indicating circular polarization (CP)
present in the signal, well above the $\leq3\%$ residual instrumental polarization effects for our data.  
For all of the events the total polarized fraction is 100\% within 
statistical errors due to thermal noise. 
CP in radio signals from CRs at the few percent level has been hypothesized
to arise from interference between the primary signal generation from geomagnetic effects~\cite{FalckeGorham, HF05},
and the secondary signal from the Askaryan effect~\cite{SLAC01},
but there is no currently accepted model to predict the resulting CP content for our signals.

 \begin{figure}[htb!]
  \includegraphics[width=3.in]{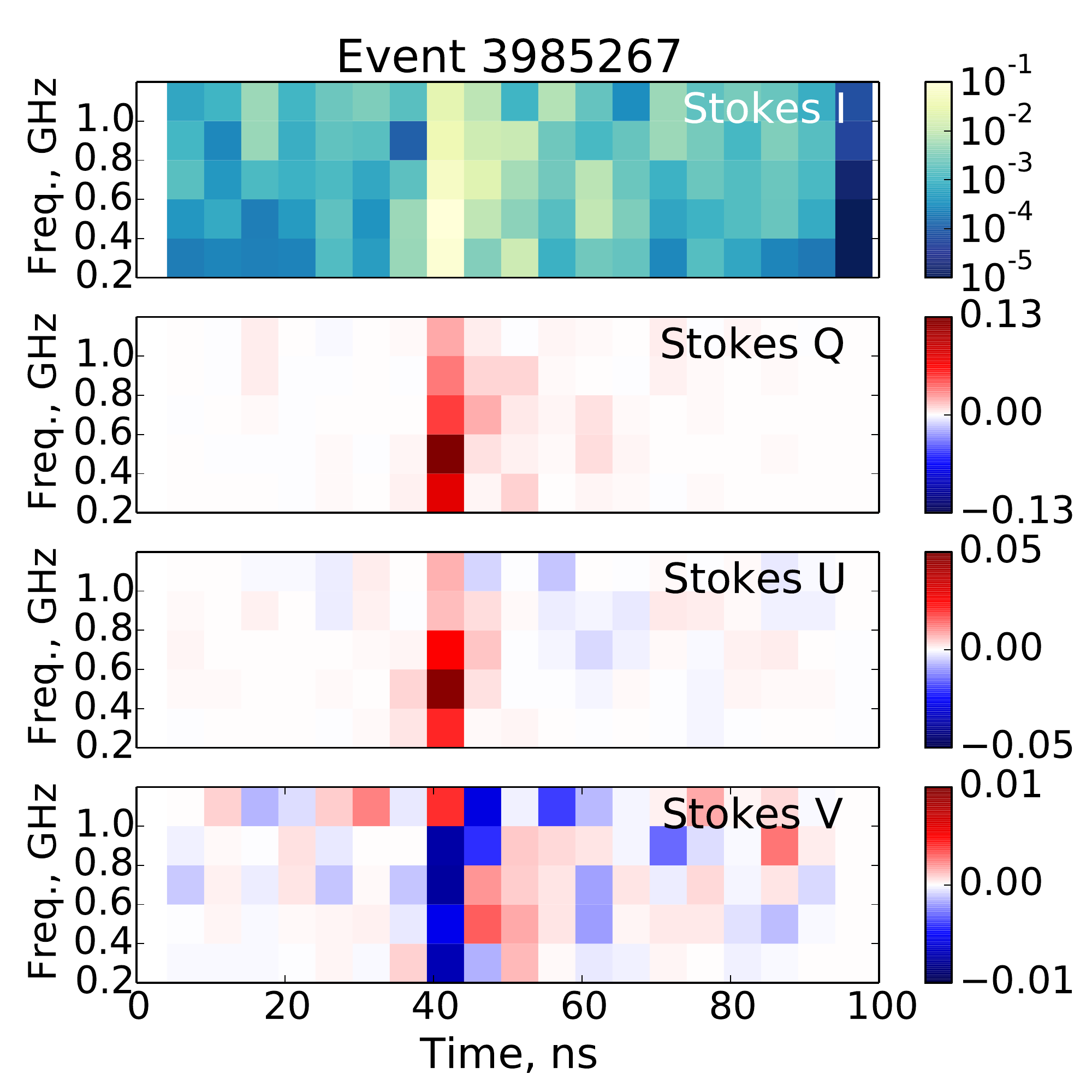}
 \caption{Stokes parameters for event 3985267.}
 \label{stokes3985}
 \end{figure}

\begin{table*}[tb!]
\caption{Parameters of event D (flight I) of unknown origin, for both the direct and reflected signal hypotheses.}
\begin{center}
\begin{threeparttable}
\begin{tabular}{ccccccccccc}
Hypothesis & Latitude &Longitude$^{\dag}$  & angle &  $D_{1200}$ &  $D_{Xmax}$  &   $H_{Xmax}$ &$D_{300}$    & $D_{100}$ &   $D_{Hmin}$  &  $H_{min}$ \\ \hline
downward CR, reflected& 83.16S & 18.9E &  $-27.4\pm0.3^{\circ}$  & 84.9 & 92.2 &  8.65 & 105.4 & 120.3& $73.1\pm0.8$& 2.59   \\
upward, direct from ice surface& 82.86S & 18.15E &  $-27.4\pm0.3^{\circ}$  & 50.7 & 62.9 &  7.0 & 69.4 & 72.0 &  $73.1\pm0.8$& 2.59   \\
upward, start 5km above ice& 82.56S & 17.4E &  $-27.4\pm0.3^{\circ}$  & * & 30.6 &  22.2 & 54 & 61 &  $63.1\pm0.8$& 7.59   \\
\hline \\
\end{tabular}
\begin{tablenotes}
\item[$^\dag$] Latitude and Longitude of the estimated location of shower maximum $X_{max}$.
\item[*] This shower exits the atmosphere at about 800 gm cm$^{-2}$ column depth.
\end{tablenotes}
\end{threeparttable}
\end{center}
\label{tbl2}
\end{table*}%

The waveform in Fig.~\ref{wfmall} for the remaining event D shows
a strong Hpol, and a correlated Vpol signal. 
The primary pulse correlates well with both the above-horizon signals and the inversion of the
14 reflected CR signals.
There is also an excess of noise evident in the trailing part of the signal, similar
to what is observed in several of the reflected CRs~\cite{HooverThesis}, although in this case
it appears more persistent and larger in amplitude.
In Fig.~\ref{stokes3985} we show the spectro-temporal plot of Stokes parameters for this event,
with clear detections of Q, U, and V, indicating both a linear and CP component;
the CP fraction is $\sim 10\%$ of the total polarization. 

Table~\ref{tbl2} shows
parameters for event D under the hypothesis that it is radio emission from a CR air shower, 
seen either in reflection from the ice surface, or from a direct air shower starting along the track from 
the surface to the payload, although for the former case the polarity is inconsistent. For the latter
case, the only Standard Model (SM) physics origin we know of for up-going air showers is from the 
interactions or decay of a secondary
lepton from a neutrino interaction; however, at these angles, the chord distance through the 
Earth most likely excludes neutrinos
of the energies that ANITA is likely to detect in such a process.

 \begin{figure}[htb!]
 \includegraphics[width=3.5in]{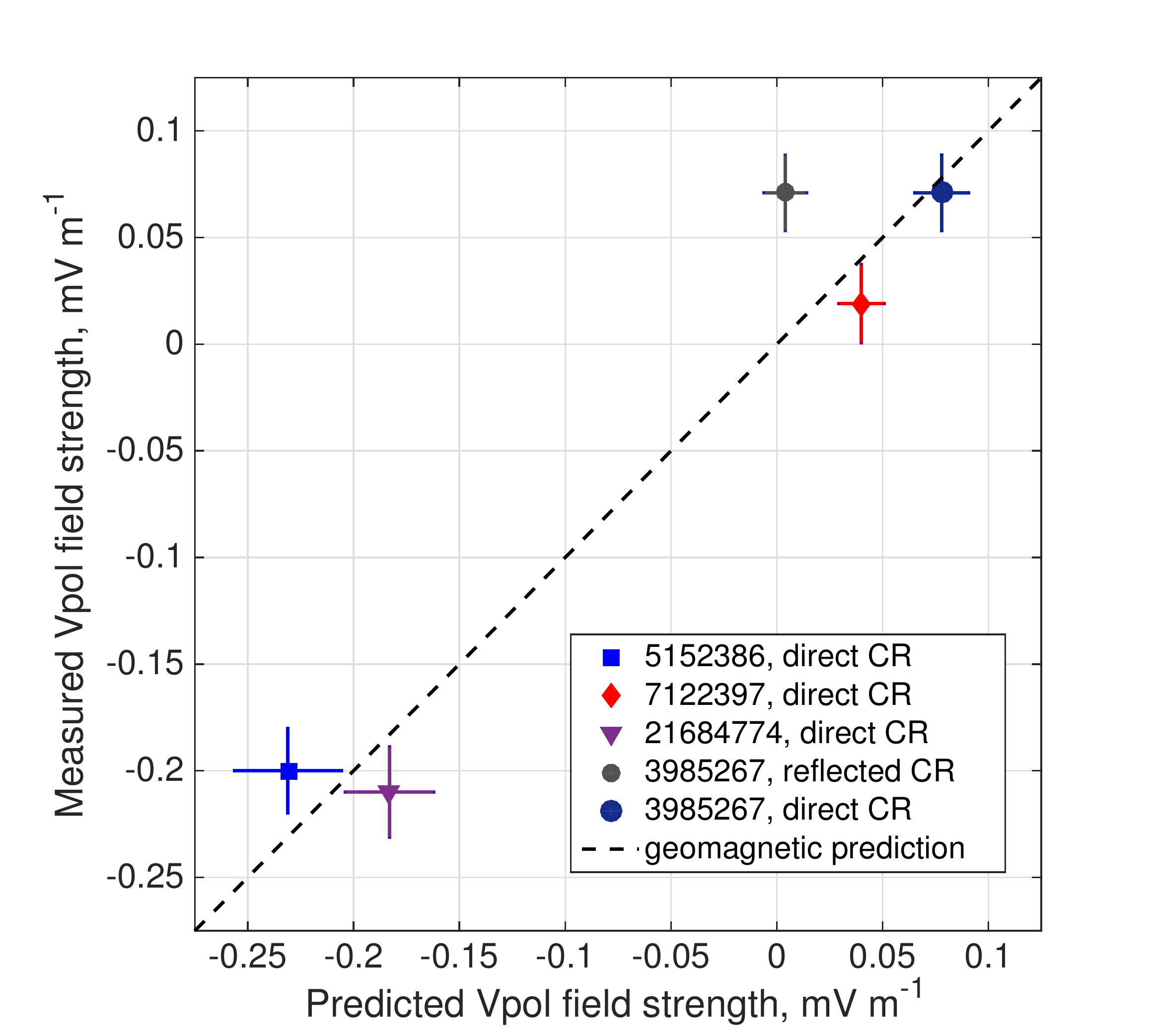}
 \caption{Geomagnetic correlation of events. The dashed line shows the prediction for
 pure geomagnetic Lorentz force-induced emission. 
 }
 \label{geomag}
 \end{figure}
 
For a cosmic-ray air shower, the Lorentz force on the relativistic electron-positron pairs
yields a plane of acceleration in the local shower frame given by $\sin \Psi = \hat{v} \times \hat{B}$,
where $\hat{v}$ is a unit vector giving the shower direction, and $\hat{B}$ the geomagnetic field
direction. The resulting radiation Poynting vector, arising primarily from the region near
shower $X_{max}$, can then be extrapolated to the payload location for each event
to determine the predicted field-strength ratio for Vpol to Hpol. 
Residual non-vertical components of the Antarctic geomagnetic fields 
will result in small but correlated Vpol components for CR events; 
anthropogenic or other backgrounds should have no correlation
to geomagnetic reference planes.
Fig.~\ref{geomag} shows results of this analysis for the four events considered here.
Errors on the predicted values arise primarily from the combined uncertainties of the Hpol field strength,
and the amplitude calibration between Vpol and Hpol. Measurement errors are dominated by
the thermal noise floor.
The stratospheric events are all consistent within errors with geomagnetic correlation,
as is the case for event D, when evaluated for the geomagnetic parameters of an
upward-coming direct event.
If the observed polarity of event D were inverted compared to what was observed, it 
could be marginally consistent with a reflected CR (at the 2.5$\sigma$ level).
However, the statistical chance of a mis-identification of the polarity is negligible, since
the coherently beam-formed signal-to-noise ratio for the field 
strength of this event is $16:1$ for Hpol, and $4:1$ for Vpol.
We thus conclude that a reflected-CR hypothesis is excluded for this event.
We note also that the measured Vpol/Hpol ratio of the largest secondary peak in this event, occuring 4~ns after
the primary peak, is consistent geomagnetically with the first peak,
suggesting a similar physical origin for these two components.

The original blind analysis that selected the ANITA-I CR events~\cite{ANITA_CR} required only
that the events show phase coherence not present in thermal noise fluctuations
and that their reconstructed position be isolated both temporally and spatially from all other events.
None of the original selection involved 
waveform correlation, correlation to geomagnetic parameters, or estimates of Stokes V content. 
For each of these independent parameters we can use the measured distributions for the background to
estimate the cumulative fraction that equal or exceed our observed values. Assuming they are
uncorrelated, the product of these individual probabilities provides an {\it a posteriori}
estimate of the probability that the background could produce this event~\cite{Suppl1}.

The fraction of the 80,000 anthropogenic background events that equal or
exceed the magnitude of event D's shape correlation coefficient with the previously identified CRs 
is $p_{wfm} = 0.022$. Anthropogenic events are uncorrelated to the Antarctic geomagnetic field, 
 and the fraction of such events that equal or exceed event D's geomagnetic correlation is $p_{geo} = 0.07$.
The fraction of events with instrumental Stokes V magnitude that exceed that of 
event D is $p_{V} = 0.05$. We estimate a trials factor of $f_{trial} = 3$ for a small number of
additional parameters investigated as potential discriminators and rejected.
Combining these factors, the estimated probability is
$p_{wfm} \times p_{geo} \times p_{V} \times f_{trial} = 2.4 \times 10^{-4}$;
Given the estimated surviving background of 1.6 events, 
we would expect $N \simeq 4 \times 10^{-4}$ possible anthropogenic 
events with characteristics like event D
in our data sample.  Anthropogenic origin for this event is thus rather strongly disfavored 
by the data.

For these three parameters, we also have measured values for our CR sample.
This allows us to form a likelihood ratio, using Bayes' theorem~\cite{Bayes},
of the CR hypothesis $CR$ to the anthropogenic hypothesis $A$:
\begin{equation}
 \frac{P(CR|E)}{P(A|E)} =  \frac{P(CR)}{P(A)} \frac{P(E|CR)}{P(E|A)} ~.
\end{equation}
where $E$ represents the experimental values.
Assuming the two hypotheses are {\it a priori} equally likely, $P(CR) = P(A)$, then
we can estimate the terms on the right directly from the data.
For the CR sample, we find $q_{wfm} = 0.13$, $q_{geo} = 0.93$,
and $q_{V} = 0.38$, where $q$ here indicates the individual probability for
event D given the CR distributions of each of the parameters noted above. 
The resulting likelihood ratio is 
$$P(CR|E)/P(A|E) = (q_{wfm} q_{geo}  q_{V}) / (p_{wfm} p_{geo} p_{V}) \simeq 550$$
where the trials factor is common to both cases.
The data thus strongly favor the CR hypothesis over the anthropogenic
hypothesis, although the latter cannot be excluded at high confidence.
This conclusion is consistent with the original analysis, which would almost certainly have
classified this event as a CR if its polarity had been inverted compared to
what was measured.

In Table~\ref{tbl3} we provide estimates of the energy of
each of the air showers considered here, based on the assumption of
scaling from simulations of down-going CR~\cite{Harm16}. 
For event D we consider only the upcoming hypothesis.
The uncertainties in each case arise primarily from
the lack of precision in the $X_{max}$ location and related systematic effects. 
More precise estimates will require detailed
simulations that are beyond our scope.
For event D, a $\tau$ decay origin
for the shower still leaves large uncertainty in the location of the decay 
along the track; indeed, a $\tau$ decay higher than about 6~km above the surface leads to
a shower that can exit the atmosphere before it even reaches shower maximum. 

\begin{table}[tb!]
\caption{Estimated energy of observed showers; uncertainties are primarily systematic.}
\begin{center}
\begin{tabular}{lcccc}
Event & (A) 5152386  & (B) 7122397 &  (C) 21684774 & (D) 3985267   \\[2pt]
\hline \\
Energy, EeV &  $9.9 \pm 3.0$ & $1.1 \pm 0.40$ & $1.2\pm 1.0$ & $0.60 \pm 0.40$  \\[2pt]
\hline
\end{tabular}
\end{center}
\label{tbl3}
\end{table}%

However, the hypothesis of a $\tau$ decay poses difficult problems of
interpretation for the parent $\nu_{\tau}$. 
The minimum emergence angle possible given our angular errors and the uncertainty of
where we are on the $\sim 1^{\circ}$ emission cone is $25.4^{\circ}$~\cite{Suppl1}, with a corresponding
chord through the Earth of 5450~km, about 20,000 km water equivalent for Earth's density profile. 
At 1~EeV, the SM neutrino interaction length is of order
$1600$~km water-equivalent, and the implied attenuation 
coefficient is $\sim 4 \times 10^{-6}$, effectively excluding a neutrino origin for this event~\cite{Suppl1}.
Regeneration of $\nu_{\tau}$~\cite{regenRefs} in the Earth can effectively reduce this coefficient by factors of
order 2-5 in some cases, but not enough to change this conclusion.
Indeed we find that, for SM cross sections, ANITA's
geometric acceptance to this type of event should lead to more events observed closer to the horizon,
which are not seen.  However, SM uncertainties can in some scenarios lead to
suppression of the $\nu$ cross section at these energies~\cite{suppressionRefs},
an important effect since it enters through the exponent of the attenuation.
Initial estimates indicate that a cross-section suppression factor of $\sim 3-5$ is required to
make this event a plausible $\nu_{\tau}$  candidate. This level of suppression
would require revision of many current UHE neutrino limits.

We note that 
the ice depth at the location of this event is 3-4~km; energy loss of
a $\tau$-lepton in ice is $\sim 1/3$  of that in crustal rock, increasing the
probability of survival to decay above the ice surface. This effect can
lead to an order-of-magnitude more acceptance for such air showers
over ice or water compared to surface land~\cite{Weiler}.
Also, the $\tau$-lepton 
may itself initiate a shower in the subsurface ice
at these high energies which may emerge with the $\tau$ and induce an early
air shower; this shower's radio emission would be delayed
relative to the higher-altitude shower produced by the $\tau$-decay.
Such a scenario could lead to the correlated trailing noise 
observed within $\leq 10$~ns of the primary peak in the
waveform of this event, as it is consistent with refractive atmospheric delay if this portion
of the signal originated near the surface of the ice.

Current or future data may be able to confirm or falsify whether neutrino interactions
are the origin of this event. To optimize detection for in-ice neutrino events, 
ANITA-II had a trigger design with
low efficiency for CR-like events~\cite{ANITA-II}. For ANITA-III's flight completed
last year, the trigger for CR events was reinstated, and data analysis is ongoing.
ANITA-IV is scheduled to fly later this year.

We thank NASA for their generous support of ANITA, and the Columbia
Scientific Balloon Facility for their excellent field support, and the National Science
Foundation for their Antarctic operations support. This work
was also supported by the US Dept. of Energy, High Energy Physics
Division.

\end{document}